\documentclass[11pt, reqno]{amsart}
\makeatletter
\g@addto@macro{\endabstract}{\@setabstract}
\newcommand{\authorfootnotes}{\renewcommand\thefootnote{\@fnsymbol\c@footnote}}%
\makeatother

\usepackage{graphicx}
\usepackage{amssymb, amsthm, amsmath}
\usepackage{amsfonts}
\usepackage{color,float,appendix}
\usepackage[all]{xy}
\usepackage{floatflt}
\usepackage{indentfirst}
\usepackage[rightcaption]{sidecap}
\usepackage[font=small,labelfont=bf, figurename=Fig.]{caption} 

\usepackage{epstopdf}
\usepackage{amsfonts}
\usepackage{setspace}
\usepackage{mathrsfs}
\usepackage{subfigure}
\usepackage{booktabs}

\makeatother

\calclayout

\numberwithin{equation}{section}

\newcommand{\bdes}{\begin{description}}
\newcommand{\edes}{\end{description}}
\newcommand{\bary}{\left \{\begin{array}{ll}}
\newcommand{\eary}{\end{array}}
\newcommand{\bmat}{\begin{bmatrix}}
\newcommand{\emat}{\end{bmatrix}}

\newcommand{\be}{\begin{equation}}
\newcommand{\ee}{\end{equation}}

\newtheorem{thm}{Theorem}[section]

\newtheorem*{ack}{Acknowledgement}
\newtheorem*{dis}{Disclosure Statement}
\newtheorem*{fund}{Funding}

\begin{document}
\begin{center}
 \LARGE
Evolutionary Dispersal of Ecological Species via Multi-Agent Deep Reinforcement Learning
     \par \bigskip

  \normalsize
 Wonhyung Choi\footnote{Email address : whchoi@kaist.ac.kr}\textsuperscript{a}
  and
  Inkyung Ahn\footnote{Corresponding Author
 Email address : ahnik@korea.ac.kr}\textsuperscript{b}
\par \bigskip
 \textsuperscript{a}
 Natural Science Research Institute, KAIST,\\ 291 Daehak-ro, Yuseong-gu, Daejeon, 34141, Republic of Korea,\\
  \textsuperscript{b}
 Department of Mathematics, Korea University,\\ 2511, Sejong-Ro, Sejong,  30019, Republic of Korea
 \par \bigskip

\end{center}

{\sc Abstract} \hspace{0.1in}
Understanding species dynamics in heterogeneous environments is essential for ecosystem studies. Traditional models assumed homogeneous habitats, but recent approaches include spatial and temporal variability, highlighting species migration. We adopt starvation-driven diffusion (SDD) models as nonlinear diffusion to describe species dispersal based on local resource conditions, showing advantages for species survival. However, accurate prediction remains challenging due to model simplifications.
This study uses multi-agent reinforcement learning (MARL) with deep Q-networks (DQN) to simulate single species and predator-prey interactions, incorporating SDD-type rewards. Our simulations reveal evolutionary dispersal strategies, providing insights into species dispersal mechanisms and validating traditional mathematical models.

\vspace{0.2in}

{\sc Key words} \hspace{0.1in} 
Evolutionary dispersal; Predator--prey interactions; Multi-agent deep reinforcement learning; Starvation-driven diffusions; Ideal free distribution

\vspace{0.2in}

{\sc AMS subject classifications:}  35J60, 35K57, 92D25, 68T05, 93E35

\section{Introduction}
Mathematical modeling and the prediction of the dynamics of ecological species with intricate interactions are crucial for comprehending ecosystems. Numerous studies have explored population models from an evolutionary dispersal standpoint, emphasizing factors such as species interactions and resource availability. Key contributions have highlighted models where dispersal is influenced by interspecies interactions \cite{Averill, Cosner2, Cosner0}. It is widely recognized that realistic dispersal models for biological species must account for both interspecies interactions and responses to environmental conditions \cite{Okubo, Skellam}.

Over the years, dispersal theory has advanced to include environmental factors as vital components \cite{Okubo}. The environment affecting a species includes a variety of elements, such as the presence of other interacting species. Typically, species migrate to areas offering more favorable habitats, with sufficient food and optimal conditions for survival. Therefore, understanding dispersal strategies is essential for studying species evolution. Detailed discussions of discrete and continuous models on dispersal evolution are available in \cite{Cantrell-Cosner, Cohen, Johnson, Nagylaki, Okubo} and the references therein.

Starvation-driven diffusion (SDD) is a representative species dispersal process characterized by nonlinear diffusion, first introduced in \cite{ck13} for a single species. This model describes species dispersal based on local habitat conditions, where resources are relatively abundant or scarce in relation to population density. The advantages of SDD for species survival have been mathematically investigated in many studies involving interacting species \cite{ck13, cca23, ca19, ca19b, ca20, cba19, kk16, kkl13, kkl14}. Furthermore, research has investigated numerous nonlinear dispersal models within this framework. Similar to earlier studies in mathematical ecology, these approaches aim to predict ecological phenomena. However, attaining more precise predictions and accounting for the intricate interactions among species continue to be challenging due to the inevitable simplifications required in mathematical analysis.

Despite the strengths of these mathematical approaches in modeling and prediction, studying population dynamics remains a complex task in computational biology. Analyzing and accurately predicting the adaptive behaviors of interacting species is intricate and often requires an agent-based approach to gain deeper insights into the subject.

The agent-based approach \cite{Holland, Macal} relies on agents that interact based on predefined rules within a simulated environment, effectively capturing both individual and collective behaviors. Unlike traditional population dynamics modeling, which typically seeks to model equilibrium mathematically, the agent-based approach does not follow this method directly. However, when combined with Monte Carlo simulations in two-dimensional space, the agent-based approach can uncover more intricate and varied spatiotemporal patterns within the studied domains. In scenarios where state transition rules and rewards are not predetermined, common in areas utilizing reinforcement learning, agents learn through environmental interactions, including responses from other agents. This interaction allows learning agents to discover and adapt to more optimal policies over time.
Recently, several studies have utilized multi-agent reinforcement learning (MARL), an agent-based approach, to analyze species behaviors \cite{plkap21, wcw19, wcw20}. Reinforcement learning (RL) agents develop strategies to maximize rewards through adaptive or learned movement in response to their habitat. This study employs deep reinforcement learning, specifically the deep Q-networks (DQN) algorithm, which has been widely applied in various fields. 

This paper investigates single species and predator-prey interactions in an ecosystem. We consider ecological species as agents that learn and evolve their dispersal strategies using MARL with appropriate rewards. The artificial environment we construct is a simplified version based on a mathematical model, which serves as a foundational step in creating more complex environments with ecological interactions and determining species' evolutionary dispersal. Specifically, we simulate species dynamics in the established ecosystem by utilizing starvation-driven diffusion type (SDD-type) rewards for MARL agents. Through this approach, we explore the dispersal mechanisms of single and interacting two species and compare the RL simulation results with the properties in the mathematical model. This research provides valuable insights into the ecological evolution of species dispersal. The comparison between the results of the MARL simulation and the mathematical model validates the mathematically modeled species dispersal. 
We point out that observations can vary between the agent-based model and the mathematical population model when a small number of agents are simulated. Therefore, in this work, we employ a large number of species agents for simulations and comparisons. This study emphasizes the importance of the agent-based perspective in understanding species dispersal mechanisms. 

The remainder of this paper is organized as follows:
Section 2 provides a brief introduction to SDD, our framework based on MARL with deep Q-network (DQN) algorithms, and the simulation environment.
Section 3 presents the simulation results for the established ecosystems, which are then compared to the mathematical results detailed in the appendix.
Finally, we summarize our results and discuss future work.

\section{Methods}
\subsection{Starvation-driven diffusion}
Before establishing the reinforcement learning of ecological species based on the starvation-driven type rewards, we first introduce \emph{starvation-driven diffusion (SDD)}, a dispersal strategy of ecological species in the mathematical model. SDD is introduced in \cite{ck13} and is based on a concept of animal behavior that the species decides to migrate to other regions depending on their habitat condition. SDD describes the way of species movement that the species migration by assuming that species dispersal adjusts in response to the relative abundance and scarcity of food resources in a region. When species assess local habitat conditions, specifically the availability of food resources, they adjust their migration speed accordingly. They move slowly in regions where food is abundant and quickly migrate to other areas when resources are scarce.

To represent such species dispersal mathematically, we define a value called \emph{a starvation measure} $s$ by (species density)/(amount of food resources) at each location, which represents the relative food scarcity. If we adopt a motility function of the species $\gamma(s)$ increasing to $s$, then the higher $s$, the higher the moving rate of the species. If we put this in the species diffusion function, $\Delta(\gamma(s)u)$ represents SDD, where $\Delta=\sum_{i=1}^N \frac{\partial^2}{\partial x_i^2}$ is a Laplace operator and $u$ is a species density function. 

One of the representative properties of the SDD model is the ideal free distribution. The ideal free distribution (IFD) theory suggests that, in a habitat with varying amounts of resources, individuals of species will distribute themselves in such a way that the number of individuals in different patches is proportional to the available resources in each patch. The distribution is considered \textit{ideal} because species make choices that maximize their access to resources, resulting in a theoretically optimal distribution for each individual in the population. In \cite{kkl14}, the authors showed that IDF is observed for the single-species model when the motility function $\gamma$ is suitably chosen and some conditions on the food resource function (Theorem~\ref{idf-math}).

The single-species model can be expanded to include multiple species with various interactions. Competition models \cite{ca19b,kk16,kkl14} have been studied, and the representative result of the SDD model is that SDD provides a species survival advantage in heterogeneous environments and enhances species fitness. In the case of the predator-prey interaction model, SDD makes the predator survive even under the condition that the predators moving randomly cannot survive (Theorem~\ref{thmone} and \ref{thmtwo}), which represents that the survival advantage of SDD is also observed in the predator-prey model. For the interested readers, we refer to some literature \cite{ca19b,kk16,kkl14} for the predator-prey SDD system with other functional responses, and detailed modeling and mathematical results for SDD are presented in the appendix.

The properties of SDD models introduced above, IDF and the survival advantage of SDD, are among the pivotal influences of evolutionary dispersal. Throughout this paper, we will observe whether the single-species model and predator-prey ecosystem established with MARL have these phenomena.

\subsection{Framework}
In this subsection, we present the framework for studying predator-prey systems through simulations by using multi-agent deep reinforcement learning. 
First, we provide a description of the environment of the predator--prey ecosystem. It is assumed that interacting predator and prey dynamics occur in a two-dimensional space consisting of $N\times N$ cells. A periodic spatial boundary condition is adopted, which indicates that when an agent passes through one side of the simulation box, it reappears on the opposite side. Each cell can be occupied by food, predators, and prey, possibly empty. To describe starvation-driven diffusion type (SDD-type) rewards, 
agents of the same type can be located simultaneously in the same cells, unlike in the existing studies \cite{plkap21, wcw19, wcw20}, which is an important difference in this paper. At the beginning of the model, the initial agents are randomly distributed in the cells, and the food resources are placed in the lattices randomly or artificially. The construction of the dynamics of agents and interactions among the agent individuals and the given resources is as follows:

(i) Prey agent: A prey agent can move to any adjacent cell and reproduce offspring in the cell at a reproduction rate that depends on how much food exists in the cell. When the predator agents are in the same cell, prey is eaten by the predators at a capture rate and dies out. Prey death also occurs by interspecific competition with a given death rate. 

(ii) Predator agent: A predator agent can move to any adjacent cell. When prey agents are in the same cell, they can reproduce offspring in the cell with a conversion rate, which is an efficiency at which predators convert consumed prey into predator reproduction. Predator death occurs by interspecific competition with a given death rate.

(iii) Rewards: Each individual agent obtains a reward through his or her action. SDD-type rewards are considered for prey and predators and are defined as follows:
\begin{align}
&\mbox{Reward of prey} = \begin{cases} \frac{\mbox{$N_{food}$}}{\mbox{$N_{prey}$ $+$ $N_{pred}$}},\quad &\mbox{if $N_{prey}$ $+$ $N_{pred}$ $\neq 0$},\\ \mbox{$N_{food}+1$},\quad &\mbox{otherwise},\end{cases} \label{preyreward}\\
&\mbox{Reward of predator} = \begin{cases} \frac{\mbox{$N_{prey}$}}{\mbox{$N_{pred}$}},\quad &\mbox{if $N_{pred}$ $\neq 0$},\\ \mbox{$N_{prey}$+1},\quad &\mbox{otherwise} , \end{cases} \label{predreward}
\end{align}
where $N_{food}$, $N_{prey}$ and $N_{pred}$ represent the number of food resources, prey agents, and predator agents in the cell, respectively, and the rewards $N_{food}+1$ and $N_{prey}+1$ are given for the case where the denominator of the reward is $0$ ($N_{prey}=0$ or $N_{pred}=0$). Here, we neglect the effect of the predator's death rate on the reward of the predator, unlike the mathematical model detailed in the appendix, because the predator's death rate does not affect learning. We point out that the SDD-type reward can be given by other types, for example, the linear combination, $N_{prey}-N_{pred}$, for the reward of the predator, but this may not represent the relativity of satisfaction as in the mathematical model.

In this work, the changes in food resources caused by consumption of prey and seasonal effects, etc., are not considered 
because we want to establish a multi-agent reinforcement learning environment similar to that of the diffusive mathematical population model for comparison with the results of the mathematical models. For the same reason, we do not consider
the age structure of species agents or the starving period when the agent individual cannot find its resources, food for prey, or prey for predators. Considering the factors we neglected could make the model complex but more realistic, which will be our future work.

\subsection{Learning with a deep Q-Network}
The reinforcement learning (RL) problem is usually formulated by the Markov decision process by the tuple $(S,A,P,r,\gamma)$, where $S$ is a set of states, $A$ is a set of actions that an agent can choose, $P$ is a state transition probability, $r$ is a reward function, and $\gamma\in[0,1)$ is a discount rate. At time $t$, when an agent takes an action $a_t\in A$ in state $s_t\in S$, the agent transitions to the next state $s_{t+1}$ with a probability $P(s_{t+1}|s_t,a_t)$, and then it obtains rewards $r_t=r(s_t,a_t)$ defined by \eqref{preyreward} and \eqref{predreward} for each type of species. Since the prey and predator are partially observable agents, the state $s_t$ will be considered as a partial state of the whole state.
As mentioned above, the prey and predator agents have nine possible actions $\{\mbox{up, down, left, right, upleft, upright, downleft, downright, remain}\}$, as shown in Figure 1. 
\begin{figure}[!htb]
\includegraphics[scale=0.2]{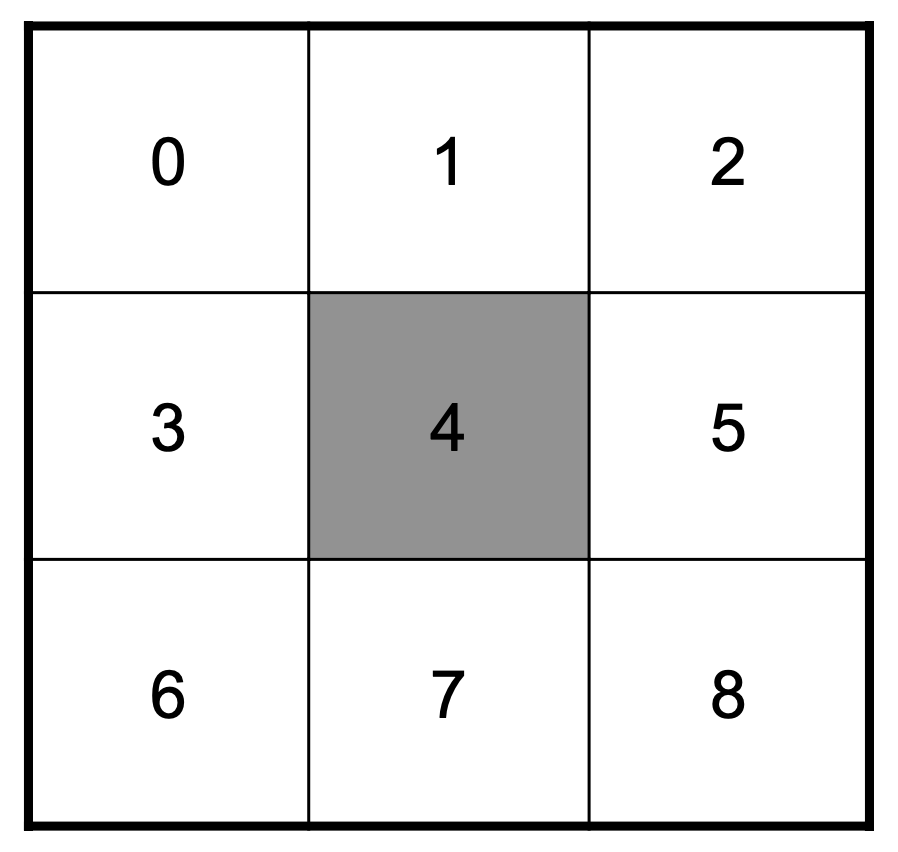}
\caption{Discrete action space}
\end{figure}
At each time step $t$, the ecological species agent determines the partially observed state $s_t$ and takes the action $a_t$ from the action space $A$ following an agent's policy, denoting policies of prey and predator $\pi_{prey}$ and $\pi_{pred}$, respectively, which maps to $S$ to $A$. The purpose of learning is to obtain the optimal policies, say $\pi_{prey}^*$ and $\pi_{pred}^*$, maximizing the expected discounted return defined by 
$$R_t:=\mathbb{E}[\sum_{k=t}^T \gamma^{k-t} r_k],$$
where $t$ is the current time step, and $T$ is the total time step at which the task is performed. For a given policy $\pi$, the action-value function is defined by $Q(s,a)= \mathbb{E}_\pi[R_t|s_t=s,a_t=a]$, and the optimal action-value function $Q^*$ satisfies the following Bellman equation with the optimal policy $\pi^*$:
\be\label{bellman}
Q^*(s,a) = \mathbb{E}[r(s,a) + \gamma \max_{a'}\sum_{s'\in S}P(s'|s,a)Q^*(s',a')],
\ee
where $s'$ and $a'$ are the next state and action, respectively. 

Q learning is an off-policy algorithm that is model-free, updating the Q-function iteratively using the Bellman equation \eqref{bellman}:
$$Q_{t+1}(s,a)= (1-\alpha)Q_t(s,a)+\alpha[r_t + \gamma \max_{a'}Q_t(s',a')],$$
where $\alpha\in[0,1]$ represents the learning rate. A deep Q-Network (DQN) approximates the action-value function, represented by $Q(s,a;\theta)$ using the weights and bias values of a neural network parameterized by $\theta$. The DQN is trained by minimizing the mean squared error (MSE) loss between the target Q-value and the predicted Q-value:
$$L(\theta)=\mathbb{E}[(r+\gamma \max_{a'}Q(s',a';\theta)-Q(s,a;\theta))^2].$$
During the training of Q-networks for prey and predator agents, an $\epsilon$-greedy exploration policy, balancing exploration and exploitation by choosing a random action with probability $\epsilon$, and the experience replay technique are employed. In this paper, we make the species learn the evolved movement using this DQN algorithm. Then, we simulate the predator--prey ecosystem with the constructed environment.

\section{Results}
\subsection{Single species model}

\begin{figure}[!htb]
\includegraphics[scale=0.5]{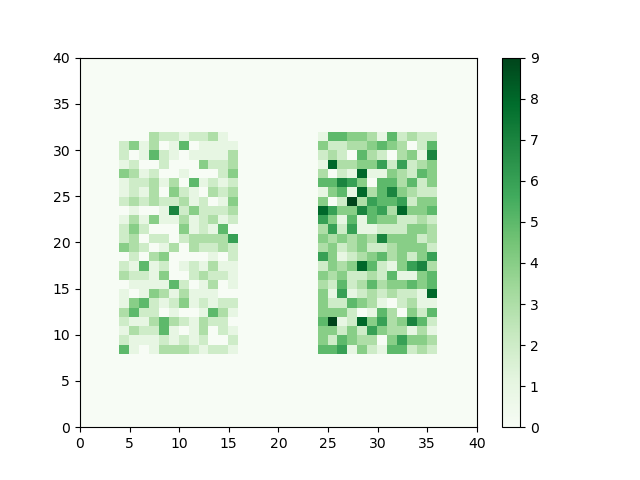}
\caption{Food resource distribution. The total amount of food resources is 1500 (500 in the left patch and 1000 in the right patch)}
\label{food_idf_dist}
\end{figure}

\begin{figure}[!htb]
\includegraphics[scale=0.3]{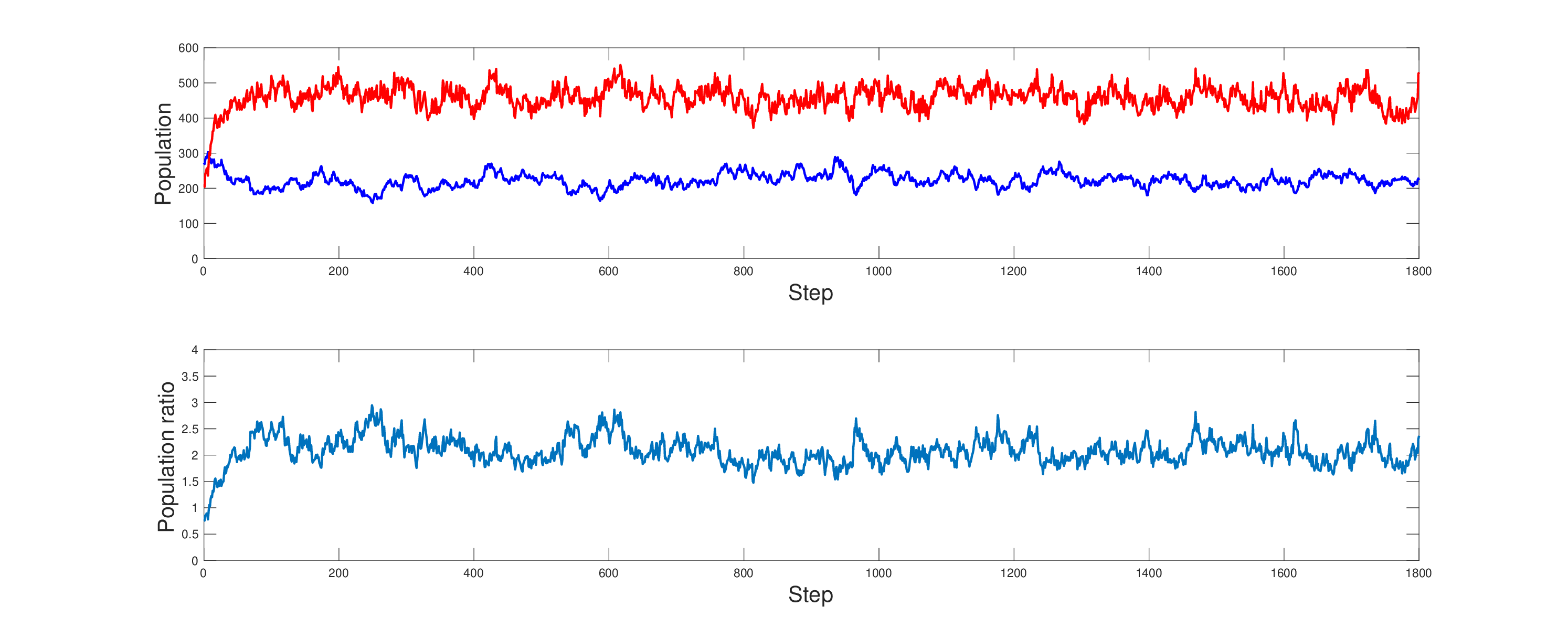}
\caption{The number of populations in each resource patch (the red line is the number of species in the right-side patch, and the blue line is the number of species in the left-side patch).}
\label{idf_rep}
\end{figure}

This subsection considers a single species with learning based on an SDD-type reward. First, the ideal-free distribution (IFD) for evolved species dispersal corresponds to the mathematical results in Section 2.1 and Appendix A.1. Recall that the ideal free distribution indicates that individuals of species will distribute themselves so that the number of individuals in different patches is proportional to the available resources in each patch. Unlike the PDE model, a simulation with MARL is performed with some randomness, and we cannot observe a stationary distribution of species density. Thus, to show the IFD of the evolved dispersal mechanism, we consider the food distribution partially in the grid given as in Fig.~\ref{food_idf_dist}. 
Adopting an SDD-type reward \eqref{preyreward} without $N_{pred}$ and density-dependent birth and death, such as the logistic type, the individuals of the species are distributed proportionally to the number of resources in each patch (Fig.~\ref{idf_rep}), which represents the evolved dispersal of species that have IFD properties.
In the simulation of Fig. ~\ref{idf_rep}, the initial total number of species is 1500 in the $40\times40$ grid. The average of population ratio between two patches is 2.0798 in this simulation (Fig. ~\ref{idf_rep}). 
Considering stochastic perturbations and the spatial heterogeneity of food resources, the system exhibits an IDF when species migrate with evolutionary dispersal.

\subsection{Predator--prey model}
\begin{figure}[!htb]
\includegraphics[scale=0.5]{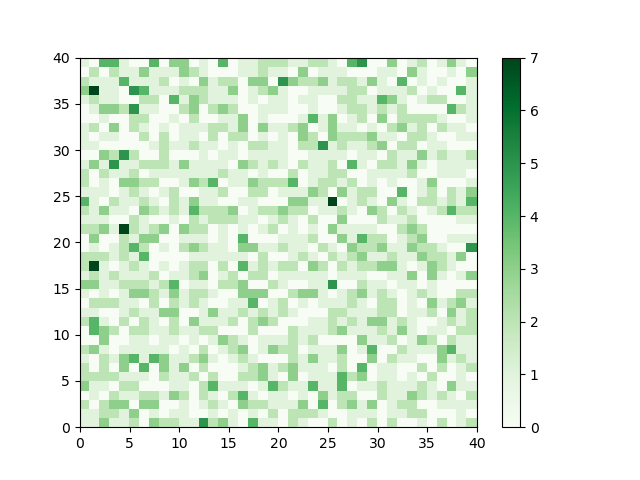}
\caption{Distribution of food resources in the grid (total amount $N_{food}=2000$).}
\label{food_dist}
\end{figure}

In this subsection, we simulate the predator--prey ecosystem for three cases: (i) both species move randomly, (ii) only prey has evolutionary dispersal, and (iii) both species have evolutionary dispersal. It will be simulated in $40\times40$ grids with parameters and initial values set as in Table~\ref{init}. 
\begin{table}[!htb]
\centering
\begin{tabular}[t]{cc}
\toprule
\bf{Notation \& Value} & \bf{Meaning}\\
\midrule
$\epsilon = 0.1$ & exploration rate in $\epsilon$-greedy policy\\
$b_{prey}=0.11$ &Reproduction rate of prey\\
$b_{pred}=0.25$ &Reproduction rate of predator\\
$d_{pred}=0.1$ &Death rate of prey\\
$d_{pred}=0.1$ &Death rate of predator\\
$n_{prey}=600$ &Initial number of preys\\
$n_{pred}=300$ &Initial number of predators\\
$N_{food}=2000$ &Number of food resources\\
\bottomrule
\end{tabular}
\caption{Parameters and initial values set used in the simulation}
\label{init}
\end{table}
The food resources are distributed randomly in the grids (Fig.~\ref{food_dist}), and the initial prey and predators are also randomly distributed in the grid. Multiple amounts of food and multiple predators and prey can be located simultaneously in the cell.

\begin{figure*}[!htb]
\centering
\subfigure[Both prey and predator move randomly]{
\includegraphics[scale=0.3]{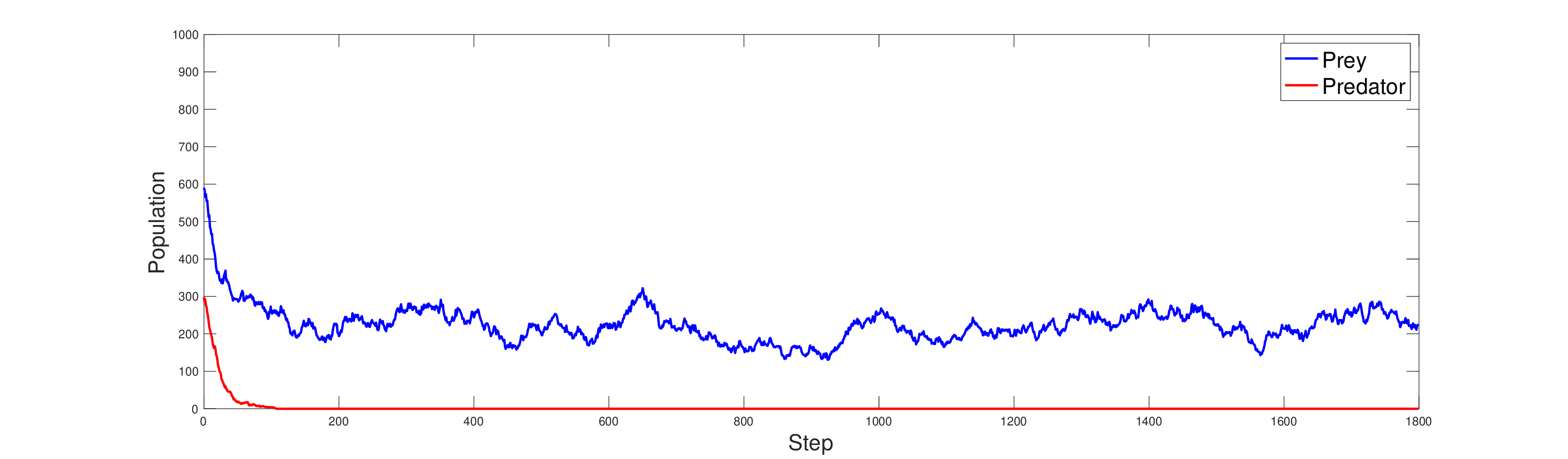}
\label{fig:nrGroup}
}
\subfigure[Prey moves with evolutionary dispersal and predators move randomly]{
\includegraphics[scale=0.3]{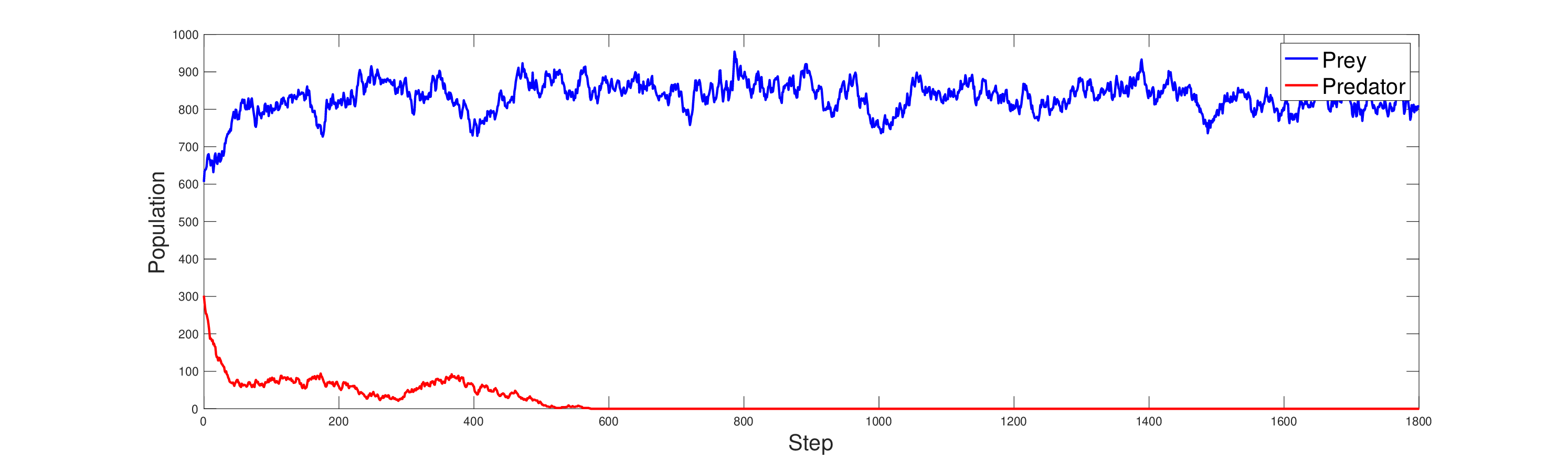}
\label{fig:overallResult}
}
\caption{Simulation results for the predator-prey ecosystem (a) without learning about both species (b) with only learning about the prey
}
\label{predex}
\end{figure*}

Recall that the mathematical results in the predator--prey system show that the evolutionary dispersal, SDD, can cause predators to survive when random diffusing predators do not survive in the given heterogeneous environment (Theorem~\ref{thmone} and \ref{thmtwo}). We first experiment with a case in which the predators move randomly. In Fig.~\ref{predex}, in the simulations for two cases, (a) the prey moves randomly, and (b) the prey moves with evolutionary dispersal, indicating that the randomly moving predator cannot survive in the system regardless of the prey dispersal type. In addition, compared with the two simulation results, we can observe that the evolutionary dispersal increases fitness, resulting in a greater number of species with evolutionary dispersal than that of randomly diffusing species, which is one of the effects of SDD and a conjectured question in the mathematical model due to the difficulty of directly comparing the two equations.

\begin{figure*}[!htb]
\centering
\subfigure{
\includegraphics[scale=0.3]{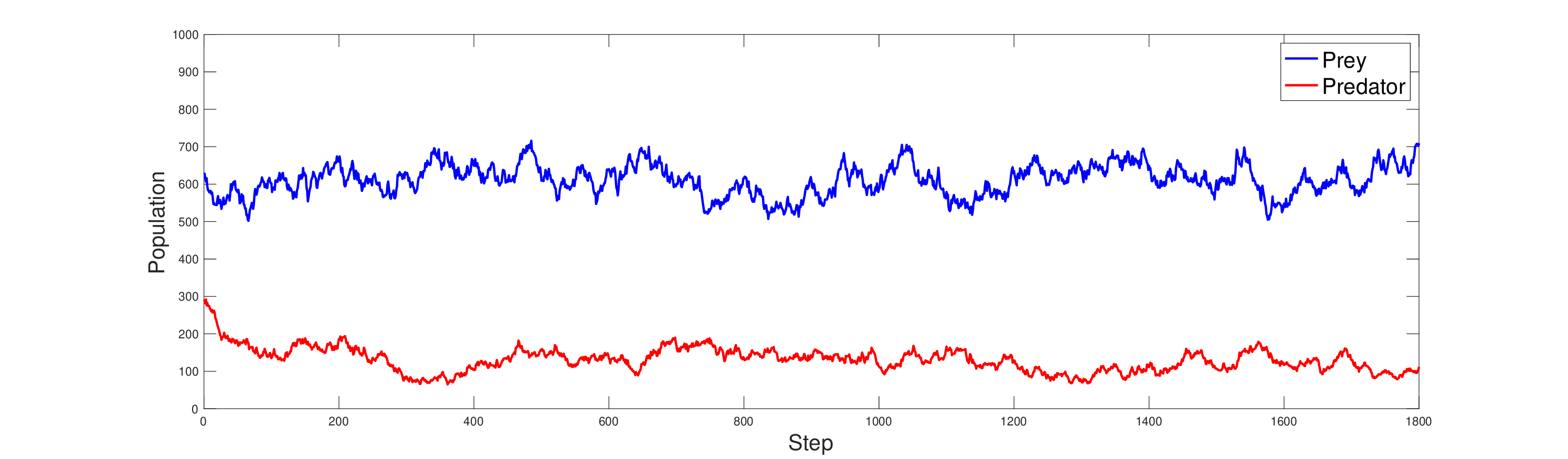}
\label{fig:nrGroup}
}
\subfigure{
\includegraphics[scale=0.3]{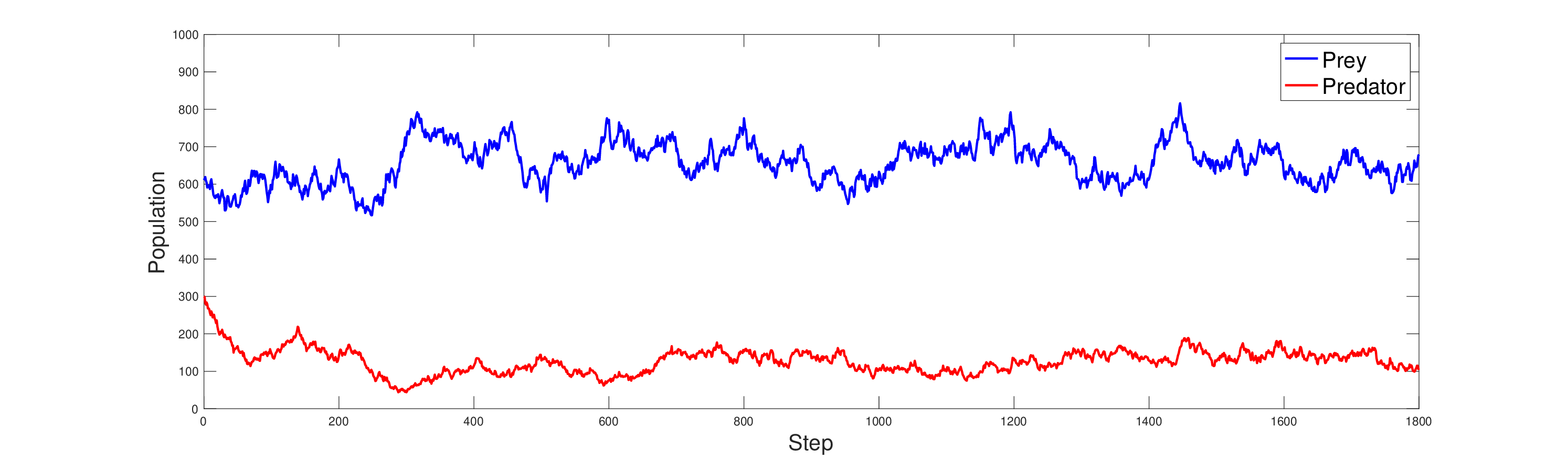}
\label{fig:overallResult}
}
\subfigure{
\includegraphics[scale=0.3]{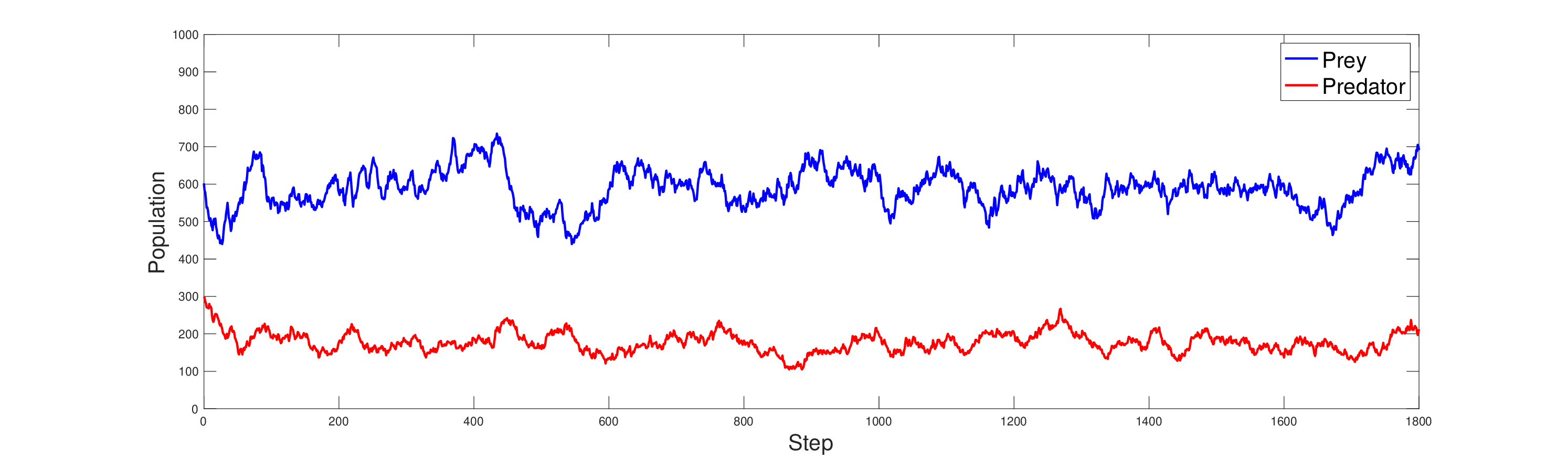}
\label{fig:overallResult}
}
\caption{Simulation results for the predator--prey ecosystem with learning for both species for three different food distributions
}
\label{coex}
\end{figure*}

Fig.~\ref{coex} shows that the result of the mathematical model also holds in the experiment in which we make both prey and predator agents learn evolutionary dispersal. When predator dispersal evolves, the predator survives under the same settings as in Table~\ref{init}, as shown in Fig.~\ref{predex}. The simulation results of coexistence and predator survival are observed in each experiment with three different randomly distributed food resources. Comparing simulation results, Fig.~\ref{predex} and ~\ref{coex} show that the evolution of predator dispersal  
 with SDD-type rewards provides a survival advantage to the species as in the mathematical properties (Theorem~\ref{thmone} and \ref{thmtwo}). For the interested readers, we remark that the same result can be obtained even if the prey moves randomly by adjusting other parameters, but we do not provide the simulation result here.

\section{Conclusion}

Our study effectively demonstrates the utility of multi-agent reinforcement learning (MARL) in simulating species dispersal and predator-prey dynamics, thereby validating theoretical models through practical application.

In the single-species model, we found that the ideal-free distribution (IFD) for evolved species dispersal corresponds well with mathematical predictions. Despite the inherent randomness in MARL simulations, the distribution of species proportionally matched the availability of resources, indicating that the evolved dispersal mechanisms exhibit IFD properties. This result confirms that species adapt their movements to maximize resource utilization, even in a stochastic environment, thereby validating the theoretical model in a more realistic setting.

In the predator-prey model, our simulations revealed that predators' evolutionary dispersal, driven by SDD-type rewards, significantly enhances their survival compared to random movement. The results showed that predators with evolutionary dispersal strategies thrive in heterogeneous environments, supporting the theoretical predictions. Furthermore, the simulations demonstrated that species with evolved dispersal mechanisms have increased fitness and better coexistence, highlighting the advantages of adaptive movement strategies in maintaining ecological balance. 

The observations, IDF and the survival advantage of predators, obtained by MARL simulation with SDD-type rewards are the representative phenomena in the reaction-diffusion models with SDD (Theorem~\ref{idf-math}-\ref{thmtwo}).

Our results suggest that MARL is one of the robust frameworks for understanding species dispersal and interactions in complex ecosystems. The strong alignment of simulation results with traditional mathematical models underscores the potential of agent-based approaches in ecological research. Future studies should explore more intricate models and diverse ecological scenarios to further validate and expand upon these findings.

\begin{ack}
\end{ack}

\begin{dis}
The authors declare no conflict of interest.
\end{dis}

\begin{fund}
This research was supported by the Basic Science Research Program through the National Research Foundation of Korea (NRF), funded by the Ministry of Science, ICT \& Future Planning (NRF-2022R1F1A1063068, RS-2024-00340022) and the Ministry of Education (NRF-2021R1A6A3A01086879). 
\end{fund}

\appendix
\appendixpage

\section{Mathematical models with SDD}
In the appendix, we present a brief introduction to starvation-driven diffusion(SDD) and some results for the mathematical model with SDD, which corresponds to the MARL simulation results in this paper.

 Cho and Kim introduced SDD for a single species \cite{ck13}, which describes the species migration by assuming that species dispersal adjusts in response to the relative abundance and scarcity of food resources in a region.
Consider a diffusive logistic population model for a single species:
\be\label{single}
u_t(x,t) = \nabla\cdot \vec{J}_u (x,t) + r u(x,t)(m(x)-u),
\ee
where $u(x,t)$ is the population density of a species at location $x\in \mathbb{R}^N$ at time $t$, $\vec{J}_u$ is the diffusion flux of  species $u(x,t)$,  $r$ is the growth rate, and $m(x)$ is a heterogeneously given food resource.

The starvation measure $s$ is defined as $\mbox{(food demand)}/\mbox{(food supply)}$. Thus, the food is relatively abundant (scarce) when $s$ is small (large). Assuming that $m(x)$ is interpreted as the food supply and that the population density $u(x,t)$ is the food demand, we have
\begin{equation}\label{sat}
s: = \frac{u(x,t)}{m(x)}.
\end{equation}
By perceiving the habitat conditions at $x\in\Omega$, such as food supplementation and the number of competing species, species choose their movement speed such that they move to other regions slowly when the location is abundant and move rapidly when they are in a region where there is not enough food. To describe this movement mechanism of species mathematically, we adopt a motility function $\gamma(s)$ that increases to $s$. Then, the SDD can be represented by nonlinear diffusion:
$$\nabla\cdot\vec{J}_u (x,t) = \Delta(\gamma(s)u)(x,t),$$
where $\Delta=\sum_{i=1}^N \frac{\partial^2}{\partial x_i^2}$ is a Laplace operator.
The Fokker-Plank type nonlinear diffusion can be written as
$$\Delta(\gamma(s)u) = \nabla\cdot\Big[ (\gamma(s)+\gamma'(s)/m)\nabla u  - \frac{\gamma'(s)}{m}u\nabla m\Big].$$
Since $\gamma' \ge 0$, we note that SDD has the effects of self-diffusion and directional movement to the food resources $m$, which describes species traits such as avoiding intraspecific competition and chasing regions with abundant food.

The starvation measure in the early works on SDD \cite{ca20,ck13,kk16,kkl13,kkl14} was defined by the reciprocal of $s$, called satisfaction measure, and the motility function $\gamma$ was assumed to decrease. 
Subsequent studies on single-species and interacting multispecies models with nonuniform dispersal have been conducted  \eqref{sat} \cite{cca23,ca19,ca19b,cba19}, and we note that the phenomena observed in earlier works can also be obtained in analogous ways. In this paper, to avoid the case where the species density is $0$ in the MARL simulation, the definition $s$ in \eqref{sat} is used, and the results that will be introduced are the equivalent versions to those in previous works.

\subsection{Ideal free distribution}
 For a partial differential equation model, the ideal free distribution(IFD) is not observed in linear diffusion, $\vec{J}_u = d\nabla u$ with a positive constant $d$. However, in \cite{kkl14}, the authors showed the optimal selection phenomenon of species dispersal with a suitably chosen form of the motility function $\gamma$ and some conditions on $m(x)$. We present the results of IFD briefly, and the detailed conditions for $\gamma$ and $m$ are not presented. We refer readers to Section 5 in \cite{kkl14} for interested readers.

 For the value $s:=u/m$, consider a motility function with $0<\ell<h$ such that
\begin{equation}\label{step}
\gamma_0(s) = \begin{cases} \ell, \quad &0\le s\le 1,\\
h,\quad &1\le s <\infty,\end{cases}
\end{equation}
and its continuous approximation 
$$\beta_\varepsilon(s) = \begin{cases} \ell, \quad &0\le s\le 1-\varepsilon,\\
h,\quad &1+\varepsilon\le s <\infty,\\
\frac{1+\varepsilon-s}{2\varepsilon}\ell + \frac{s-(1-\varepsilon)}{2\varepsilon}h,\quad&\mbox{otherwise}.
\end{cases}$$
Then, we can approximate the discontinuous motility function $\gamma_0$ by
$$\gamma_\varepsilon(s)=\beta_\varepsilon * \eta_{\varepsilon^2}(s)=\int_{\mathbb{R}}{\beta_\varepsilon(y)\eta_{\varepsilon^2}(s-y)dy},$$
where $\eta_{\varepsilon^2}$ is a symmetric mollifier supported in $(-\varepsilon^2,\varepsilon^2)$. 

Let $u_\varepsilon(t,x)$ be a solution of 
\begin{equation}\label{single}
\begin{cases}
u_t = \Delta(\gamma_\varepsilon(s)u) + u [m(x)-u],\quad &(t,x)\in[0,\infty)\times\Omega,\\
\nabla(\gamma_\varepsilon(s)u)\cdot \vec{n}=0,\quad &(t,x)\in[0,\infty)\times\partial \Omega,
\end{cases}
\end{equation}
where $\vec{n}$ is a unit outward normal vector, and the boundary condition is given by the Neumann boundary condition, i.e., no flux on the boundary $\partial\Omega$. For small $\varepsilon>0$, \eqref{single} has a unique positive steady-state solution $\theta_\gamma$, which is globally asymptotically stable \cite{kkl14}, i.e., $u_\varepsilon(t,x)\to \theta_\gamma(x)$ as $t\to\infty$. Note from \cite{kkl14} that \eqref{single} with for a general increasing function $\gamma$ also has a globally asymptotically stable unique positive steady-state solution under a certain assumption. Then, the following result is obtained for the ideal free distribution in \cite{kkl14}. 
\begin{thm}\label{idf-math}
Let $\theta_\gamma$ be the solution of \eqref{single}. Suppose that $m(x)$ satisfies
$$\ell \max_{\bar{\Omega}}m(x)<\frac{h+\ell}{2}\min_{\bar{\Omega}}m(x)<\frac{h+\ell}{2}\max_{\bar{\Omega}}m(x)<h\min_{\bar{\Omega}}m(x).$$
Then, $\theta_\gamma(x)$ converges to $m(x)$ as $\varepsilon\to0$.
\end{thm}
The assumption in Theorem~\ref{idf-math} is satisfied when the difference between $\ell$ and $h$ is large, which means that species react and decide to move sensitively to the environment.

\subsection{Predator-prey model}
 In this subsection, we present some results for the following
  predator--prey model:
\begin{equation}\label{pp}
\begin{cases}
u_t = \Delta(\gamma_{\varepsilon,1}(s_1)u) + u [m(x)-u] -\alpha uv\\
v_t = \Delta(\gamma_{\varepsilon,1}(s_2)v) + \beta uv - (c+dv)v,\quad &(t,x)\in[0,\infty)\times\Omega,\\
\nabla(\gamma_{\varepsilon,1}(s_1)u)\cdot \vec{n}=\nabla(\gamma_{\varepsilon,2}(s_2)u)\cdot \vec{n}=0,\quad &(t,x)\in[0,\infty)\times\partial \Omega,
\end{cases}
\end{equation}
where $u$ and $v$ are the population densities of the prey and predator, respectively. The parameters $\alpha,\beta,c$ and $d$ are positive constants, representing the capturing rate by the predator, the conversion rate, the predator's intrinsic death rate, and the intraspecific competition rate of the predators, respectively. For simplification, we assume that all parameters are given by $1$ except for $c$. 

Unlike the single species model \eqref{single}, the satisfaction measures for the prey and predator are dependent not only on the food resources but also on the interacting species. Thus, we define the measures $s_1$ and $s_2$ as follows:
\begin{equation}
s_1 = \frac{u}{m+v},\quad s_2 = \frac{c+v}{u}.
\end{equation}
The motility functions $\gamma_{\varepsilon,i}$ for $i=1,2$ are defined by
$\gamma_{\varepsilon,i} = \gamma_{0,i} * \eta_\varepsilon,$
where $\gamma_{0,i}$ is a form of \eqref{step} with $\ell = \ell_i$ and $h=h_i$, and a mollifier $\eta_\varepsilon$ is supported in $(-\varepsilon,0)$. Throughout this subsection, we assume that $\varepsilon$ is sufficiently small. 

We note that the predator cannot survive without prey in a model \eqref{pp}, and the model has only one semitrivial solution $(u_0,0)$ exists, where $u_0$ satisfies
\begin{equation}\label{pp}
\begin{cases}
\Delta(\gamma_{\varepsilon,1}(s)u_0) + u_0 [m(x)-u_0]=0,\quad &x\in\Omega,\\
\nabla(\gamma_{\varepsilon,1}(s)u_0)\cdot \vec{n}=0,\quad &x\in\partial \Omega,
\end{cases}
\end{equation}
with $\tilde{s}=u_0/m$. For the predator--prey model \eqref{pp}, the possibility of predator survival depends on the stability of $(u_0,0)$, and the stability is determined by the sign of the principal eigenvalue
$$\lambda_1=\sup_{\phi\not\equiv 0, \phi\in W^{1,2}(\Omega)}\frac{\int_\Omega{-|\nabla\phi|^2 + \frac{-c+u_0}{\gamma_{\varepsilon,2}(s')}\phi^2 dx}}{\int_\Omega{\phi^2 dx}},$$
where $s'=u_0/c$. The detailed procedures for obtaining the linearized eigenvalue problem and the following stability results
 are described in \cite{ca20}. 

\begin{thm}\label{thmone} 
Let $(u_0,0)$ be the semitrivial solution of \eqref{pp}. For fixed $\ell_2$ and $h_2$, there exists $c^{*}(\gamma)\in(0,\max_{\bar{\Omega}}u_0)$ depending on $\Omega, \ell_2$ and $h_2$ such that \\
(i) if $c<c^{*}(\gamma_{\varepsilon,2})$, then $(u_0,0)$ is  
unstable, and\\
(ii) if $c>c^{*}(\gamma_{\varepsilon,2})$, then $(u_0,0)$ is locally asymptotically stable.\\
In particular, $c^{*}(\gamma_{\varepsilon,2})>c^*(\ell_2)$.
\end{thm}
Next, we define the value $M$ by
\begin{equation} \label{M}
M=\frac{\int_{\{u_0<c\}}{c-u_0}}{\int_{\{u_0>c\}}{u_0-c}},
\end{equation}
which is the degree of the predator's environment with respect to the difference in the available prey and the predator's death rate. The value is well defined when $u_0>c$ somewhere in $\Omega$. Let us denote the average of $u_0$ by $\bar{u}_0$. Then, we have the following result:
\begin{thm}\label{thmtwo} 
Let $(u_0,0)$ be the semitrivial solution of \eqref{pp}. \\
(i) If $c\geq \max_{\bar{\Omega}}u_0$, then $(u_0,0)$ is locally asymptotically stable for any $\ell_2$ and $h_2$.\\
(ii) If $c<\bar{\theta}_\mu$, then $(u_0,0)$ is unstable for any $\ell_2$ and $h_2$.\\
(iii) Suppose that $c\in(\bar{u}_0,\max_{\bar{\Omega}}u_0)$. 
There exists $M>1$ such that
 if $\frac{h_2}{\ell_2}>M$, then $(\bar{u}_0,0)$ is locally asymptotically unstable. \\
(iv) Suppose that $c\in(\bar{u}_0,\max_{\bar{\Omega}}u_0)$ and $\frac{h_2}{\ell_2}<M$.
There exists $\ell^*(\gamma_{\varepsilon,2})>0$ such that\\
if $\ell_2<\ell^*(\gamma_{\varepsilon,2})$, then 
 $(u_0,0)$ is 
  unstable; and\\
if $\ell_2>\ell^*(\gamma_{\varepsilon,2})$, then 
 $(u_0,0)$ is locally asymptotically stable.\\
Moreover, $\ell^*(\gamma_{\varepsilon,2})>\ell^*(\ell_2)$.
\end{thm}
The relations in Theorem~\ref{thmone} and \ref{thmtwo}, $c^{*}(\gamma_{\varepsilon,2})>c^*(\ell_2)$ and $\ell^*(\gamma_{\varepsilon,2})>\ell^*(\ell_2)$, indicate the effect of starvation-driven diffusion on predator survival. For $c\in (c^*(\ell_2),c^{*}(\gamma_{\varepsilon,2}))$, Theorem~\ref{thmone} shows that a predator obeying evolutionary dispersal can survive, but a randomly diffusing predator cannot survive under the same conditions, which means that SDD provides the species with a survival advantage. The same result is derived from Theorem~\ref{thmtwo} for $\ell_2\in (\ell^*(\ell_2),\ell^{*}(\gamma_{\varepsilon,2}))$. Furthermore, the greater the sensitivity of SDD, $\frac{h_2}{\ell_2}$, is, the greater the probability of predator survival (Theorem~\ref{thmtwo}-(iii)). 

\end{document}